# Management of Social and Economic Development of Municipalities

Maria A. Shishanina[1], Anatoly A. Sidorov[1]
[1]Tomsk State University of Control Systems and Radioelectronics
Tomsk, Russian Federation
mariia.a.shishanina@tusur.ru

*Abstract* — The paper discusses the process of social and economic development of municipalities. A conclusion is made that developing an adequate model of social and economic development using conventional approaches presents a considerable challenge. It is proposed to use semantic modeling to represent the social and economic development of municipalities, and cognitive mapping to identify the set of connections that occur among indicators and that have a direct impact on social and economic development.

## I. Introduction

The territorial reality of the Russian Federation is such that management of social and economic development (SED) of regions and individual municipalities cannot rely on solutions that are considered standard in the world practice. The unique aspects of development of municipalities determine an essential role of regional and municipal authorities in addressing the problems of social and economic development of territories. However, the problem is exacerbated by the fact that, after the market transformations that the country has experienced, state and municipal authorities were not fully prepared to build an effective region-municipality relationship, which is explained by various regulatory and legal circumstances, financing and the lack of a comprehensive system of information and analytical support that would make it possible to forecast development of territories. In that context, scientific justification of decision-making in municipal SED is a high-priority objective regardless of the development level of any specific territories.

## II. Considerations of Social and Economic Development Management in Municipalities

As established by the current legislation (Federal Law No. 172-FZ "On Strategic Planning in the Russian Federation" as of June 28, 2014), regional and municipal SED is based on strategic management that is represented as a single process of development and implementation of fundamental long-term and mid-term policies designed to achieve certain goals in the context of a changing external and internal environment [1-3]. Therefore, municipal SED management represents a continuous process of developing, approving and implementing managerial decisions that includes progress monitoring, strategy development, plan implementation, and evaluation of programs. Figure 1 offers a general view of the algorithm of strategic management

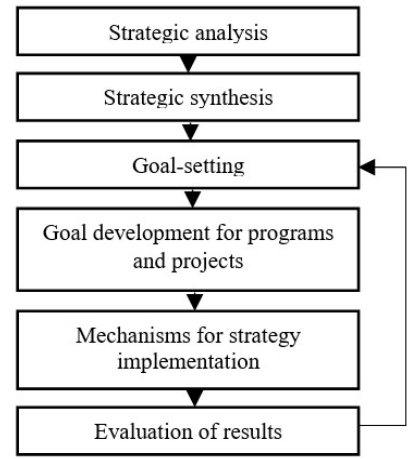

Fig.1. Algorithm of strategic management.

Analysis of fundamental documents that deal with strategic SED shows that the process of SED management can be represented as a model (figure 2) that demonstrates the decisive role of state administration in planning.

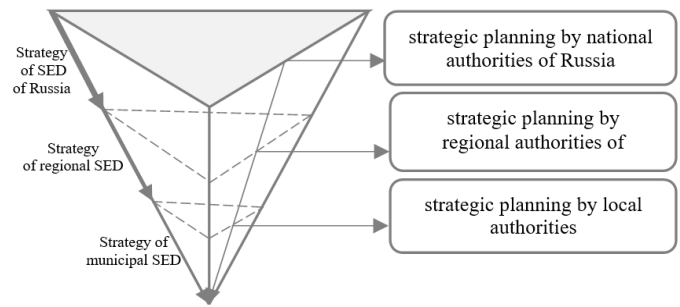

Fig. 2. Top-down SED management.

Ideally, municipal strategic management should be based on strategic plans of local governments that take into account the public opinion and opinions of the business community. However, in practice, the situation is different. Local authorities try to implement the strategy based on the existing administrative methods. As a result, all activities are reduced to an action plan that exists for executive authorities only, excluding the local community, business, etc. Another negative factor is the shortage of staff in municipalities, since their employees demonstrate a skill gap in strategic and project management, and any changes in the external or internal environment only aggravate the problems with implementation of the SED strategy. As a result, the following key problems of municipal-level strategizing can be identified [5-6]:
- non-systemic structure of the strategic development process;
- predominant reliance on administrative methods of management;
- insufficient methodological support of local initiatives from the regional-level government;
- skill gap in strategic and project management in public officials at various levels (especially in municipalities);
- heavy financial dependence, since municipalities usually have no additional sources of income;
- lack of an effective mechanism for collaboration of local authorities with the local community, business, and other stakeholders in the development.

Given that the system of strategic planning at the municipal level requires a revision, which is almost impossible to do at the level of the municipality, excluding the federal level, there is an objective need to improve the scientific and methodological justification of managerial decisions made by regulatory and administrative authorities in the process of planning, forecasting and assessment of comprehensive social and economic development of a territory. As a result, the SED management model (top-down) is transformed into a bottom-up model (figure 3) that will make it possible to take into account the unique aspects of individual territories when making long-term development plans in compliance with the national strategic goals [7].

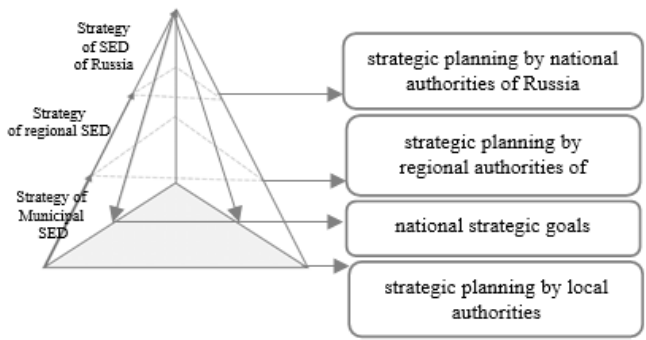

Fig. 3. Bottom-up SED management.

Despite the fact that the model shown in Figure 2 is conceptually different from Figure 3, it can be implemented in the context of the current regulatory, financial, labor, and other constraints. Management of the SED of territories (and more specifically, of municipalities) poses a sophisticated challenge that requires a creative approach, but at the same time, the process is regulated by the legislator (with an established list of required documents, recommended methods for situation analysis, etc.), which can eventually reduce any creativity to a set of templates. However, there is no common template that could be reasonably applied to all municipalities in Russia (or in a specific region), because each territory is unique and has its own development features. At the same time, it can be assumed that across the entire territory of Russia, there are municipalities that have a comparable level and direction of their development (e.g., climate, population, specialization, etc.). Such municipalities will serve as model units that will be used to develop template strategies. This approach will make it possible to develop a certain level of template support for the process of planning and management of SED. As a result, the existing system of SED will be based both on regional goals and on municipal initiatives that will vary depending on the types of territories. Thus, in order to achieve national goals and improve the efficiency of territory management at all levels, it is important to consider the unique aspects of development of specific municipalities and represent SED management in a comprehensive way, as a semistructured system [8-10].

III. CONCLUSION

SED of a territory is a complex and continuous process of planning and forecasting that engages authorities of all levels. However, despite the formalized approach to planning at the federal level, municipalities face a number of problems in the process of this work. The greatest challenge in the process of decision-making in SED is the unavailability of complete information about the territory to the individual decision-maker. That is due to the fact that SED is a semistructured subject area, which fact must be taken into account in the process of planning and forecasting. To neutralize the negative impact of problems identified during the study, the authors propose to use semantic and cognitive tools in practical management of municipal SED, which will ultimately improve the quality and level of justification of managerial decisions made by officials.

ACKNOWLEDGMENT

This paper is designed as part of the state assignment of the Ministry of Science and Higher Education; project FEWM-2020-0036.